\documentclass[12pt]{article}
\RequirePackage[margin=1in,top=1in,bottom=1in]{geometry} 

\usepackage{setspace}
\usepackage{verbatim}
\usepackage[font=small]{caption}
\usepackage{multirow}
\usepackage{amsmath,amsthm,amssymb,url}
\usepackage{graphicx}
\newtheorem{prop}{Proposition}

\newtheorem*{remark}{Remark}
\usepackage{color}
\usepackage{natbib}
\usepackage{prodint}
\usepackage{mathtools}

\newcommand{\be}{\begin{eqnarray*}}
\newcommand{\ee}{\end{eqnarray*}}
\newcommand{\ben}{\begin{eqnarray}}
\newcommand{\een}{\end{eqnarray}}

\renewcommand{\d}{\mathrm{d}}

\usepackage{authblk}
\author[1]{Robert L. Strawderman\footnote{Corresponding author: robert\_strawderman@urmc.rochester.edu}}
\author[1,2]{Benjamin R. Baer}
\affil[1]{Department of Biostatistics and Computational Biology, University of Rochester, Rochester, NY, USA, 14642}
\affil[2]{School of Mathematics and Statistics, University of St Andrews, St Andrews, Fife, Scotland, KY16 9SS}

\date{}

\title{On the Role of Volterra Integral Equations in Self-Consistent,
Product-Limit, Inverse Probability of 
Censoring Weighted, and Redistribution-to-the-Right Estimators
for the Survival Function}

\begin{document}
\singlespacing
\maketitle

\begin{abstract}
This paper reconsiders several results of historical and current importance to nonparametric estimation of the survival distribution for failure in the presence of right-censored observation times, demonstrating in particular how Volterra integral equations of the first kind help inter-connect the resulting estimators.  The paper begins by considering Efron's self-consistency equation, introduced in a seminal 1967 Berkeley symposium paper.  Novel insights provided in the current work include the observations that (i) the self-consistency equation leads directly to an anticipating Volterra integral equation of the first kind whose solution is given by a product-limit estimator for the censoring survival function; (ii) a definition used in this argument immediately establishes the familiar product-limit estimator for the failure survival function; (iii) the usual Volterra integral equation for the product-limit estimator of the failure survival function leads to an immediate and simple proof that it can be represented as an inverse probability of censoring weighted estimator (i.e., under appropriate conditions). Finally, we show that the resulting 
inverse probability of censoring weighted estimators, attributed to a highly influential 1992 paper of Robins and Rotnitzky, were implicitly introduced in Efron's 1967 paper in its development of the redistribution-to-the-right algorithm. All results 
developed herein allow for ties between failure and/or censored observations.

\end{abstract}

\newpage

\section{Introduction}

\citet{Efron1967} introduces the concept of self-consistency to motivate an
iteratively specified estimator for the left-continuous version of
the survival function under random right censoring. Assuming that a solution to this iterative scheme exists, it is shown to be unique and satisfy a recursive equation \citep[Thm.\ 7.1]{Efron1967}. In a key corollary to this result, \citeauthor[Cor.\ 7.1]{Efron1967} shows that the left-continuous version of the Kaplan-Meier estimator \citep{KM1958} coincides with the self-consistent estimator over a range defined by the observed data.  Proof by induction is used to establish both results. \citeauthor{Efron1967} then introduces the redistribution-to-the-right (RTTR) algorithm as an alternative way to understand the self-consistent (hence Kaplan-Meier) estimator, demonstrating how the result produced by running this algorithm equals the self-consistent estimator. 

\citet{Satten2001}, motivated by developments in \citet{RR1992}, later prove that the Kaplan-Meier estimator for the failure distribution function can be represented as an inverse-probability of censoring weighted (IPCW) estimator in which the denominator uses a Kaplan-Meier estimator
for the censoring survival function. They remark that this result is ``somewhat unsatisfactory'' due to the use of the latter in deriving this equivalence. A constructive proof using ``coupled'' inverse-probability weighted estimators is then given that avoids using the Kaplan-Meier estimator of the censoring survival function.

The overarching intent of this work is to show how these results are easily established, and directly interconnected, through the Volterra theory of product integration as formulated in \citet{Gill1990} and summarized in \citet[Sec.\ II.6]{ABGK1993}. In particular, after reviewing  the self-consistency equation for the survival function for failure in Section \ref{sec:results:SC-review}, the following results are established
in Section \ref{sec:results:SC-VI}: (i) it is shown that Efron's self-consistency equation directly generates a Volterra integral equation for the survival function of the {\em censoring} distribution, the solution being given by its corresponding product-limit estimator; then, using this result, (ii) a transparent non-inductive proof is given that shows the product-limit estimator of the survival function for failure, hence Kaplan-Meier estimator, solves Efron's self-consistency equation. Of particular interest in (i) is our use of the {\em anticipating} form of Volterra integral equation \citep[e.g.,][p.\ 1526]{Gill1990}, as it allows us to establish the indicated result for general failure and censoring distributions. In Section \ref{sec:results:IPCW}, we continue studying the product-limit estimator, where (iii) it is shown that the IPCW estimator for the cumulative distribution function for failure  \citep[e.g.,][]{Satten2001} follows easily from the corresponding Volterra integral equation that yields the product-limit estimator for the survival function for failure; and, (iv) a necessary and sufficient condition is given for the latter to also be represented as an IPCW estimator. Finally, Section \ref{sec: RTTRA} reconsiders Efron's RTTR algorithm; there, it is shown that the resulting estimator at convergence leads directly to an IPCW estimator for the failure survival function, arguably being the earliest such example of IPCW estimation.

This paper makes several pedagogically novel contributions 
in the areas of self-consistency, product-limit estimation, and IPCW estimation. Indeed, although the results that the product-limit estimator solves Efron's self-consistency equation and that the product-limit estimator can be represented as an IPCW estimator are both well-known, the approach taken to establish these results is new, and elegantly ties all known non-parametric estimators for the failure survivor function together in a transparent and interesting way.  

\section{Main Results}

\label{sec:results:volterra}

\subsection{A brief review of Efron's self-consistency equation}
\label{sec:results:SC-review}
Suppose $T > 0$ and $U > 0$ are independent random variables having respective survival functions $S(t) = P\{ T > t\}$ and $K(t) = P\{ U > t\}$ for $t \geq 0$. Define $X = \min\{T,U\}$ and $D = I\{ T \leq U\},$ and assume that the available data takes the form of an independent and identically distributed sample $(X_i,D_i), i = 1 \ldots n.$
Define 
\begin{equation*}
 \phi_{S(\cdot)}(t; X,D) := 
I\{ X > t \} + (1-D) I\{ X \leq t\} \frac{S(t)}{S(X)},
\end{equation*}
as a random function of $t$; under the usual convention that 0/0 = 0, to be assumed throughout, this function is  well-defined for $t \geq 0.$ 
Assuming that $P\{ S(U) > 0 \} > 0$, an easy calculation also shows that 
\begin{equation}
\label{eq:needs-assp}
    \phi_{S(\cdot)}(t; X,D) = P [ T>t \mid X, D]
\end{equation}
and hence that $E[ \phi_{S(\cdot)}(t; X,D) ] = S(t)$. This last identity motivates the ``self-consistent'' estimator for $S(t)$ as originally proposed in \citet{Efron1967}. 

Specifically, define for $t \geq 0$ the naive estimator $\widehat{S}_0(t)  := n^{-1} \sum_{i=1}^n I\{ X_i > t \}.$
Let ${\cal D}_0 = \{ t \geq 0: \widehat{S}_0(t) > 0 \}$ and assume that $t \in {\cal D}_0.$ Then, for $k \geq 1$, one can recursively define the sequence of estimators
\begin{equation}
\label{eq:it-sch}
    \widehat{S}_k(t) := \frac{1}{n} \sum_{i=1}^n 
\phi_{\widehat{S}_{k-1}(\cdot)}(t; X_i,D_i).
\end{equation}
A limit of this sequence exists as $k \rightarrow \infty$ \citep{rogers1980less}; call this limit $\widehat{S}(t)$. Since $\widehat{S}_0(t) > 0,$ $\widehat{S}(t)$ is positive and satisfies the fixed-point identity, or self-consistency equation,
\begin{equation}
\label{SC1}
    \widehat{S}(t) 
    = \widehat{S}_0(t) + \frac{1}{n} \sum_{i=1}^n
    (1-D_i) I\{ X_i \leq t\} \frac{\widehat{S}(t)}{\widehat{S}(X_i)}. 
\end{equation}

The estimator $\widehat{S}(t)$ solving \eqref{SC1} above has been defined for $t \in {\cal D}_0;$ this set includes $t = 0,$ where \eqref{SC1} implies the trivial solution $\widehat S(0) = 1.$ Technically, equation \eqref{SC1} also remains well defined for $t > 0$ such that $\widehat{S}_0(t)= 0$ under the convention that $0/0 = 0$. This is consistent with the iterative scheme in \eqref{eq:it-sch}, which trivially stays at zero when it is started at zero. It may be noted, however, that the derivation of \eqref{eq:needs-assp} that motivates \eqref{eq:it-sch}, hence \eqref{SC1}, itself requires that $P(S(U)>0) > 0.$ Consequently, the fact that \eqref{SC1} also characterizes the solution when $\widehat{S}(t) = 0$ stems primarily from \eqref{eq:it-sch} and the indicated convention. For later reference, we define $\widehat{S}_{\mathit{SC}}(\cdot)$ as the self-consistent estimator for $t \geq 0$; in particular,  $\widehat{S}_{\mathit{SC}}(t) = \widehat{S}(t),$ the non-trivial solution to \eqref{SC1} for $t \in {\cal D}_0,$ and $\widehat S_{\mathit{SC}}(t) = 0$ for $t \in {\cal D}_0^C = \{t \geq 0: \widehat S_0(t) = 0\}.$
 
\begin{remark}
Although the development of self-consistent estimation above is for the right-continuous function $S(t)$, \citet{Efron1967} actually introduces the concept of self-consistency in the context of estimating its left-continuous version $S(t-) = P(T \geq t)$. In particular, in the absence of ties between failures and/or censored observations, \citeauthor{Efron1967} defines the self-consistent estimator as the solution to
\[
    \widehat{S}(t-) 
    = \widehat{S}_0(t-) + \frac{1}{n} \sum_{i=1}^n (1-D_i) I\{ X_i < t\} \frac{\widehat{S}(t-)}{\widehat{S}(X_i-)}. 
\]
\citeauthor{Efron1967} argues that $\widehat{S}(t-) = 0$ for $t > X_{(n)} = \max_{i=1,\ldots,n} X_i$, and proves that the solution coincides with the left-continuous version of the Kaplan-Meier estimator for $S(t-)$ when $t \leq X_{(n)}$ \citep[Cor.\ 7.1]{Efron1967}. 
These results imply that the solution to \eqref{SC1} agrees with the Kaplan-Meier estimator for $t < X_{(n)},$ a fact
proved using different tools, and in greater generality, in the next section.
\end{remark}

\subsection{Efron's self-consistency equation generates a Volterra integral equation having a product-limit estimator as its solution}

\label{sec:results:SC-VI}

In this section, we characterize the solution to \eqref{SC1} using  results from the theory of Volterra integral equations \citep{Gill1990, ABGK1993}. In the process, we also establish  a transparent connection between the self-consistent and product-limit estimators. 

To allow for the possibility of ties, we assume  that there exists $m$ unique observed survival times $X_i$ such that $0< X_{(1)} < X_{(2)} < \cdots < X_{(m)}$, where $m \leq n$. It will also be convenient in the coming developments to employ counting process notation.
For $i = 1 \ldots n$ and $t \geq 0$, let $N_i(t) = I\{ X_i \leq t, D_i = 1\}$, $C_i(t) = I\{ X_i \leq t, D_i = 0\}$, and $Y_i(t) = I\{ X_i\geq t\}$. 
Let $N(t) = \sum_{i=1}^n N_i(t)$, $C(t) = \sum_{i=1}^n C_i(t)$, and $Y(t) = \sum_{i=1}^n Y_i(t);$ in addition, define $\Delta N(t) = N(t) - N(t-)$ and $\Delta C(t) = C(t) - C(t-)$ for every $t > 0$. Finally, let $Y^{\dagger}_i(t) = I\{ X_i\geq t, D_i = 0 \text{ or } X_i > t, D_i = 1 \}$, and define $Y^{\dagger}(t) = \sum_{i=1}^n Y^{\dagger}_i(t)$.  

The modified at-risk process $Y^{\dagger}_i(t)$ was recently studied by \citet{baer2023recurrent}. It is easily seen that $Y^{\dagger}_i(t) = Y_i(t) 
- I\{ X_i = t, \Delta_i = 1\};$ hence, $Y^{\dagger}(t) = Y(t) - \Delta N(t).$ When $\Delta N(t) = 0$, $Y^{\dagger}(t) = Y(t),$ but not otherwise. This highlights the fact $Y(t)$ and $Y^{\dagger}(t)$ are respectively the proper choice of at-risk processes for failure and censoring when ties between failure and censoring times can exist. The reason these processes are not necessarily equal stems from the asymmetric nature of $D = I\{ T \leq U\},$ which classifies any instance of $T=U$ as a failure.  As will become clear below, the process $Y^{\dagger}(t)$ specifically reflects the fact that one is no longer considered to be at risk for censoring when a failure occurs at the same time. 

Equation \eqref{SC1} trivially implies $\widehat S(0) = 1$ and that $\widehat S(t) = 0$ on ${\cal D}_0^c = \{t: \widehat S_0(t) = 0 \}.$ Since $\widehat S_0(t) = n^{-1} Y(t+)$ and $Y(t+) = 0$ for $t \geq X_{(m)},$ it therefore suffices to characterize the solution to the self-consistency equation \eqref{SC1} for $t \in (0,X_{(m)}).$ To do so, and using the aforedefined notation, we first rewrite \eqref{SC1} for such $t$ as
\begin{equation}
\label{SC2}
    \widehat{S}(t) 
    = \frac{Y(t+)}{n} + \frac{\widehat{S}(t)}{n} \int_{(0,t]} \frac{\d C(u)}{\widehat{S}(u)},
\end{equation}
where the integral on the right-hand side is a Lebesgue-Stieltjes integral \citep[e.g.,][]{carter2012lebesgue}.  Any solution $\widehat{S}(t)$ to \eqref{SC2} is necessarily non-increasing; moreover, since $t < X_{(m)}$ implies that $Y(t+) > 0$, it must also satisfy $\widehat S(t) > 0$ there. It follows that the integral on the right-hand side is well-defined since $C(u)$ is a right-continuous step-function and $\widehat S(u) > 0$ for $u \leq t$. 

Because $\widehat S(t) > 0$ for $t < X_{(m)},$ we can divide \eqref{SC2} through by $\widehat S(t)$ to obtain
\begin{equation}
\label{SC2a}
    1 
    = \frac{Y(t+)}{n \widehat{S}(t)} + \frac{1}{n} \int_{(0,t]} \frac{\d C(u)}{\widehat{S}(u)}.
\end{equation}
Next, define for $t < X_{(m)}$ the function
\begin{equation}
\label{Khat}
    \widehat{K}(t)   
    := \frac{Y(t+)}{n \widehat{S}(t)},
\end{equation}
where $\widehat{S}(\cdot)$ is defined 
through \eqref{SC2}.
We stress here that \eqref{Khat} is merely 
a definition; that is, the notation on 
the left-hand side is being defined through the expression on the right-hand side, the latter depending directly on the (unknown) solution $\widehat{S}(t)$ that solves \eqref{SC2}. Using this definition and equation \eqref{SC2a},
equation  \eqref{SC2} can now be rewritten
   \begin{equation}
    \label{VoltK}
        \widehat{K}(t) 
        = 1 -  \int_{(0,t]} \widehat{K}(u) \frac{\d C(u)}{Y(u+)}.
    \end{equation}
The integral on the right-hand side is again a Lebesgue-Stieltjes integral, and is well-defined since $Y(u+)$ and $C(u)$ are right-continuous monotone functions with left-hand limits and $1/Y(u+)$ is positive for $u \leq t < X_{(m)}.$ Importantly, the resulting expression \eqref{VoltK} is an example of an \emph{anticipating} Volterra integral equation \citep{helton1977two,helton1978solution}; see also \citet[p.\ 526]{Gill1990} for a brief review. Using this fact, the proposition below characterizes the solution to \eqref{VoltK} as a product integral \citep{Gill1990}.

\begin{prop}
\label{prop:K2}
    Under \eqref{Khat}, $\widehat{K}(0) = 1;$
    moreover, $\widehat{K}(t)$ satisfies the anticipating Volterra integral equation \eqref{VoltK}
    for $0 < t < X_{(m)}.$ For such $t,$ 
    equation
    \eqref{VoltK} has the unique right-continuous solution
    \begin{equation}
        \label{PL-gen-K}
        \widehat{K}(t) 
        = \Prodi_{(0, t]} \left\{ 1 - \frac{\d C(u)}{Y^{\dagger}(u)} \right\},
    \end{equation}
    where $Y^{\dagger}(u) = \sum_{i=1}^n I\{ X_i \geq u, D_i = 0 \text{ or } X_i > u, D_i = 1 \}$
and $\prodi$ denotes the product integral.
Moreover, if $\Delta N(u) \Delta C(u) = 0$ for $u <  X_{(m)}$, then
\eqref{PL-gen-K} reduces to
    \begin{equation}
\label{K-PL}
    \widehat{K}(t) 
    = \Prodi_{(0, t]} \left\{ 1 - \frac{\d C(u)}{Y(u)} \right\}
\end{equation}
and $\widehat{K}(t)$ necessarily also satisfies the non-anticipating  Volterra integral equation
\[
\widehat{K}(t)  = 1 -  \int_{(0,t]} \widehat{K}(u-) \frac{\d C(u)}{Y(u)}.
\]
\end{prop}
The estimators \eqref{PL-gen-K} and \eqref{K-PL} are product-limit estimators for the censoring survival function. The latter arises as a special case of \eqref{PL-gen-K} when $\Delta N(u) \Delta C(u) = 0$ for $u < X_{(m)};$ that is, under a ``no common discontinuities'' condition (e.g., such as when $m=n$, where ties of any kind are absent). However, when this condition fails, only \eqref{PL-gen-K}, which employs the censoring at-risk process $Y^{\dagger}(\cdot)$ is a suitable estimator for the censoring survival function $K(t) = P(U > t)$ \citep[cf.\ ][p.\ 36]{Gill1980}; see also \citet[Lemma 2]{baer2023recurrent}.

\begin{remark}
The assumption that $\Delta N(u) \Delta C(u) = 0$ 
for $u < X_{(m)}$ holds almost surely when at least one of $T$ and $U$ are absolutely continuous random variables. More generally, while $\{ N_i(u) - N_i(u-) \} \{ C_i(u) - C_i(u-) \} = 0$ for $u \geq 0$ and every $i$ by construction, this does not imply $\Delta N(u) \Delta C(u) = 0;$ counterexamples are easily constructed when $T$ and $U$ are each discrete with common support.
\end{remark}

\begin{remark}
A commonly recommended procedure for estimating $K(\cdot)$ is to (i) swap $D_i$ with $1-D_i$ for all observations in a data set (i.e., so that observed censored observations become ``failures'' and vice versa); then, (ii) calculate the product-limit estimator for ``failure'' as defined in this modified dataset. Doing so returns $\widehat{K}(t)$ as defined in \eqref{K-PL}, which only consistently estimates $K(t)$ under independent censoring provided $\Delta N(u) \Delta C(u) = 0$ for $u \in [0,t]$.  
\end{remark}

The results above show that the self-consistency equation for $\widehat{S}(t)$ directly generates a Volterra integral equation for $\widehat{K}(t)$ that has the product-limit estimator  \eqref{PL-gen-K} as its solution (i.e., an estimator of the censoring survival function). This observation is interesting for at least two reasons: (i) it highlights a fundamentally important connection between estimating the failure and censoring survival functions; and, (ii) equation \eqref{SC2} itself does not generate an obvious Volterra integral equation for $\widehat S(t).$ Combining Proposition \ref{prop:K2} with the definition of $\widehat K(t)$ in \eqref{Khat} yields the unique estimator for $S(t)$ that solves the self-consistency equation.

\begin{prop}
\label{prop:PL-KM}
The product-limit estimator
    \begin{equation}
    \widehat{S}_{\mathit{PL}}(t) :=
\Prodi_{(0,t]} \left\{ 1 - \frac{\d N(u)}{Y(u)} \right\} 
\label{eq:fail-prod-lim}
\end{equation}   
is the unique solution to 
\eqref{SC2} for $0 < t < X_{(m)}.$
For $t \geq 0,$ it follows that the self-consistent estimator $\widehat S_{\mathit{SC}}(t)$ may be written
        \begin{equation}
        \widehat{S}_{\mathit{SC}}(t) =
        \widehat{S}_{\mathit{PL}}(t) I\{ t < X_{(m)} \}.
         \label{eq:fail-prod-lim-SC}
    \end{equation}
\end{prop}

A version of Proposition \ref{prop:PL-KM} was originally proved in \citet{Efron1967} using inductive arguments assuming all observed survival times are unique (i.e., no ties). The current result is more general and is obtained as a consequence of existing theory for Volterra integral equations. This novel derivation highlights the important connections between the product-limit estimators for the survival and censoring distributions \citep[cf.\ ][]{Gill1980}.
 
\begin{remark}
Consider $t < X_{(m)}.$ Substitution of \eqref{eq:fail-prod-lim}
into the right-hand side of \eqref{Khat} necessarily results in the general  form of $\widehat{K}(\cdot)$ given in \eqref{PL-gen-K}.
In practice, this means that $K(\cdot)$ is easily estimated using \eqref{Khat} once $\widehat S_{\mathit{PL}}(t)$ in \eqref{eq:fail-prod-lim} is computed,  avoiding the need for data manipulation like that described in the previous remark.
\end{remark}

As pointed out in \citet[p. 260]{ABGK1993}, the behavior of commonly used nonparametric estimators for $S(t)$ typically differ only in the extreme right tail.  The self-consistent estimator $\widehat S_{\mathit{SC}}(t)$ in \eqref{eq:fail-prod-lim-SC} equals the product-limit estimator $\widehat S_{\mathit{PL}}(t)$ in \eqref{eq:fail-prod-lim} for $t < X_{(m)}.$ However, the estimator $\widehat S_{\mathit{PL}}(t)$ is itself well-defined for all $t > 0$ under the convention that 0/0 = 0 \citep{Gill1980, ABGK1993}. Using this fact, it can be easily shown that $\widehat S_{\mathit{PL}}(t) = \widehat S_{\mathit{SC}}(t) = 0$ for $t \geq X_{(m)}$ when $\Delta N(X_{(m)}) = Y(X_{(m)})$ (i.e., all observations with $X_i = X_{(m)}$ have $D_i = 1$, hence are failures). However, when $\Delta N(X_{(m)}) < Y(X_{(m)}),$ the estimator $\widehat S_{\mathit{PL}}(t) > 0$ remains constant for $t \geq X_{(m)},$ exceeding $\widehat S_{\mathit{SC}}(t) = 0$ there.
Finally, as originally defined, the Kaplan-Meier estimator $\widehat S_{KM}(t)$ agrees with both $\widehat S_{\mathit{PL}}(t)$ and $\widehat S_{\mathit{SC}}(t)$ for $t \geq 0$ when $\Delta N(X_{(m)}) = Y(X_{(m)}).$ However, when $\Delta N(X_{(m)}) < Y(X_{(m)}),$ we have $\widehat S_{KM}(t) = \widehat S_{\mathit{PL}}(t) = \widehat S_{\mathit{SC}}(t)$ only for $t < X_{(m)};$ Kaplan and Meier leave $\widehat S_{KM}(t)$ undefined for $t \geq X_{(m)}$ beyond assuming that it respects the natural range restriction $\widehat S_{KM}(t) \in [0,\widehat S_{\mathit{PL}}(X_{(m)})]$ \citep[cf.\ ][p. 463]{KM1958}. 
Succinctly, it follows that  $\widehat S_{\mathit{SC}}(t) \leq \widehat S_{KM}(t) \leq \widehat S_{\mathit{PL}}(t)$ for $t \geq 0,$ with equality holding for $t < X_{(m)},$ and also for $t \geq X_{(m)}$ when $\Delta N(X_{(m)}) = Y(X_{(m)}).$

\subsection{The product-limit estimator for $S(t)$ solves a Volterra integral equation that generates an IPCW estimator}
\label{sec:results:IPCW}

Theorems II.6.1 and II.6.6 in \citet[Sec.\ II.6]{ABGK1993} imply that $\widehat{S}_{\mathit{PL}}(t)$ in \eqref{eq:fail-prod-lim} uniquely solves the (non-anticipating) Volterra integral equation 
\begin{equation}
\label{VoltS}
    \widehat{S}_{\mathit{PL}}(t) 
    = 1 - \int_{(0,t]} \widehat{S}_{\mathit{PL}}(u-) \frac{\d N(u)}{Y(u)}
\end{equation} 
for all $t > 0$. 
Below, we establish how and under which conditions \eqref{VoltS} leads to natural IPCW estimators of the cumulative distribution function and the survival function of $T$.

Define the IPCW estimator \citep[cf.\ ][]{RR1992}
\begin{equation}
\label{eq:ipcw-defs}
    \widehat{F}_{\mathit{IPCW}}(t) := \frac{1}{n} \sum_{i=1}^n \frac{D_i I \{X_i \leq t\} }{\widehat{K}(X_i-)} 
 \end{equation}
with domain $t>0$, where $\widehat{K}(\cdot)$ is defined in \eqref{PL-gen-K}. This estimator is 
is well-defined since $\widehat{K}(\cdot)$ is non-increasing and $\widehat{K}(X_{(m)}-) > 0.$ 
A version of this IPCW estimator of $F(t)  = P \{ T \leq t \}$ using $\widehat{K}(\cdot)$ as defined in \eqref{K-PL}  was studied in \cite{Satten2001}, who also note the need for its replacement with $\widehat{K}(\cdot)$ in \eqref{PL-gen-K} when ties can exist. 
Below, we show that \eqref{VoltS} leads directly to $\widehat{F}_{\mathit{IPCW}}(t);$ we then consider the problem of directly estimating $S(t)$ using  IPCW and its connections to both $\widehat{F}_{\mathit{IPCW}}(t)$ and $\widehat{S}_{\mathit{PL}}(t)$.

Proceeding informally for the moment, the key relation \eqref{Khat} that defined $\widehat{K}(\cdot)$ together with results summarized in previous propositions implies that $\widehat{K}(u-) \widehat{S}_{\mathit{PL}}(u-) = Y(u)/n$ and hence that $\widehat{S}_{\mathit{PL}}(u-) = n^{-1} Y(u) / \widehat{K}(u-)$ for $u \leq X_{(m)}$. Similarly to $\widehat{S}_{\mathit{PL}}(\cdot),$ it is easy to see that the product-limit estimator $\widehat{K}(t)$ in \eqref{PL-gen-K} is also well-defined for $t \geq X_{(m)};$ under the convention that 0/0 = 0, inspection of the tail behavior of  both $\widehat{K}(\cdot)$ and $\widehat{S}_{\mathit{PL}}(\cdot)$ then shows that  $\widehat{S}_{\mathit{PL}}(u-) = n^{-1} Y(u) / \widehat{K}(u-)$ for $u > 0$. Substitution of this expression for $\widehat{S}_{\mathit{PL}}(u-)$ into the right-hand side \eqref{VoltS} now gives the relationship  
    \begin{equation*}
        1 - \widehat{S}_{\mathit{PL}}(t) 
        = \frac{1}{n}  \int_{(0,t]}  \frac{\d N(u)}{\widehat{K}(u-)}.
    \end{equation*}
Simplifying the resulting integral leads to the desired result, formalized below.
\begin{prop}
\label{prop:cdf-ipw}
    Let $t \geq 0$ and define $\widehat F_{\mathit{PL}}(t) := 1-\widehat{S}_{\mathit{PL}}(t)$ 
    as an estimator for the distribution function of $T$, 
    where $\widehat{S}_{\mathit{PL}}(t)$ satisfies \eqref{VoltS} for $t > 0$. 
    Then, $\widehat{F}_{\mathit{PL}}(t)
        = \widehat{F}_{\mathit{IPCW}}(t).$
\end{prop}

That is, the product-limit estimator and IPCW estimators for $F(t) = 1-S(t)$ agree for $t \geq 0.$ Since $\widehat{S}_{\mathit{PL}}(t) = 1 - \widehat F_{\mathit{PL}}(t)$ by definition, the result of this proposition further implies that $\widehat{S}_{\mathit{PL}}(t) =  1 - \widehat{F}_{\mathit{IPCW}}(t);$ that is, 
\begin{equation}
    \widehat{S}_{\mathit{PL}}(t) = 1 - \widehat{F}_{\mathit{IPCW}}(t)
    = 1 - \frac{1}{n}  \sum_{i=1}^n \frac{D_i  }{\widehat{K}(X_i-)} + \widetilde{S}_{\mathit{IPCW}}(t),
    \label{bigres}
\end{equation}
where
\[
\widetilde{S}_{\mathit{IPCW}}(t) := \frac{1}{n} \sum_{i=1}^n \frac{D_i I \{X_i > t\} }{\widehat{K}(X_i-)}.
\]

The IPCW estimator $\widetilde{S}_{\mathit{IPCW}}(t)$ is an obvious analog to $\widehat{F}_{\mathit{IPCW}}(t)$ for estimating $S(t),$ and it is also frequently equated to $\widehat{S}_{\mathit{PL}}(t)$ in the literature. However, the identity in \eqref{bigres} clearly shows that $\widetilde{S}_{\mathit{IPCW}}(t) = 1-\widehat{F}_{\mathit{IPCW}}(t),$ hence $\widetilde{S}_{\mathit{IPCW}}(t) = \widehat{S}_{\mathit{PL}}(t),$ if and only if
\begin{equation}
        \frac{1}{n}  \sum_{i=1}^n \frac{D_i  }{\widehat{K}(X_i-)}
        = 1.
    \label{eq:keysum2}
\end{equation}
The following result gives a useful identity involving the summation appearing on the left-hand side of  \eqref{eq:keysum2}, and leads to a a simple necessary and sufficient condition for \eqref{eq:keysum2}, hence for $\widetilde{S}_{\mathit{IPCW}}(t) = 1-\widehat{F}_{\mathit{IPCW}}(t),$ to hold.
\begin{prop}
\label{thm:surv-ipw}
    The following identity holds:
    \begin{equation}
     \frac{1}{n}  \sum_{i=1}^n \frac{D_i  }{\widehat{K}(X_i-)}
     + \frac{\Delta C(X_{(m)})}{n \widehat{K}(X_{(m)}-)} = 1.
    \label{eq:keysum}
    \end{equation}
    Therefore, \eqref{eq:keysum2} holds if and only if $\Delta N(X_{(m)}) = Y(X_{(m)})$
    (i.e., $\Delta C(X_{(m)}) = 0$). Consequently, $\widetilde{S}_{\mathit{IPCW}}(t) = 
    1- \widehat{F}_{\mathit{IPCW}}(t)$ for $t \geq 0$ if and only if $\Delta N(X_{(m)}) = Y(X_{(m)})$.
\end{prop}

The condition that $\Delta N(X_{(m)}) = Y(X_{(m)})$ is equivalent to requiring that all observations with $X_i = X_{(m)}$ have $D_i = 1$; this means that all observations occurring at the last observation time are failure times and results in $ \widetilde{S}_{\mathit{IPCW}}(t) = \widehat{S}_{\mathit{PL}}(t) = 1- \widehat{F}_{\mathit{IPCW}}(t) = 0$ for $t \geq X_{(m)}$. Under this same condition, it is not difficult to show that $\widehat{K}(t) > 0$ for $t \geq X_{(m)}$. Finally, when the conditions specified in Proposition \ref{thm:surv-ipw} do hold, the IPCW estimator $\widetilde{S}_{\mathit{IPCW}}(t)$ agrees with  $\widehat{S}_{\mathit{SC}}(t)$ for $t \geq 0.$ However, when $\Delta N(X_{(m)}) < Y(X_{(m)})$, the IPCW estimator $\widetilde{S}_{\mathit{IPCW}}(t)$ does not agree with any of the other estimators, at least in finite samples.

\subsection{The RTTR Estimator is an IPCW Estimator}
\label{sec: RTTRA}

In addition to $\widehat{S}_{\mathit{SC}}(t-),$ \citet[p. 842-3]{Efron1967} introduced the redistribute-to-the-right (RTTR) estimator, defined as the limit of the RTTR algorithm. Under the setting where $m = n$ (i.e., no ties), \citet{Efron1967} argues that the RTTR estimator for $S(t-) = P\{ T \geq t \},$ say $\widehat S_{\mathit{RTTR}}(t-),$ equals $\widehat{S}_{\mathit{SC}}(t-)$ for $t > 0.$
The key to this result lies in equation (7.19) of \citet[p. 843]{Efron1967}, where it is further argued that $\widehat S_{\mathit{RTTR}}(t-) - \widehat S_{\mathit{RTTR}}(t) = (7.19)$ when (i) $t$ is a failure time or (ii) $t= X_{(n)}$ regardless of failure or censoring status; otherwise, $\widehat S_{\mathit{RTTR}}(t-) - \widehat S_{\mathit{RTTR}}(t) = 0.$
Written in terms of our notation, and for the case where $m = n$ so that the ordered observation times are given by $X_{(1)} < \cdots < X_{(n)},$ one can write
\begin{equation}
\label{eq:jumpval}
\widehat S_{\mathit{RTTR}}(X_{(k)}-) - \widehat S_{\mathit{RTTR}}(X_{(k)}) =
 J_{(k)} D_{(k)} I\{ k < n \} + 
J_{(n)} I\{ k = n \} 
\end{equation}
where $D_{(k)}$ is the failure indicator associated with $X_{(k)}$ and 
\begin{equation}
        J_{(k)} 
        = \frac{1}{n} \prod_{j=1}^{k-1} \left\{ 1 + \frac{1}{n-j} \right\}^{1-D_{(j)}}
        =\frac{1}{n} \prod_{j:X_j< X_{(k)}}
        \left\{ 1 + \frac{\Delta C(X_j)}{Y(X_j+)} \right\}.
        \label{eq:J-Ef}
\end{equation}
Notice the special status given to the largest observation in \eqref{eq:jumpval}: this observation contributes to a jump in $\widehat S_{\mathit{RTTR}}(\cdot)$ regardless of failure status. Considering the last expression on the right-hand side of \eqref{eq:J-Ef}, it can be seen that
\begin{equation}
        J_{(k)} 
        = \frac{1}{n} \Prodi_{(0, X_{(k)})} \left\{ 1 + \frac{\d C(u)}{Y(u+)} \right\}
        =\frac{1}{n \widehat K(X_{(k)}-)},
        \label{eq:J}
    \end{equation}
where the last equality follows immediately from earlier calculations; see, in particular, the proof of Proposition \ref{prop:K2}. 
To the authors' knowledge, the earliest paper to recognize the connection between the increments \eqref{eq:jumpval} and IPCW weighting appears to be \citet{malani95}; 

When ties can exist between failure and/or censoring times, the increments \eqref{eq:jumpval}  require modification. In particular, if $m \leq n$ and $X_{(1)} < \cdots < X_{(m)}$ are the unique observation times, a natural extension of Efron's arguments leading
to his equation (7.19) gives
\begin{equation}
\label{eq:jumpval2}
\widehat S_{\mathit{RTTR}}(X_{(k)}-) - \widehat S_{\mathit{RTTR}}(X_{(k)}) =
 J_{(k)} \Delta N(X_{(k)}) I\{ k < m \} + 
J_{(m)} Y(X_{(m)}) I\{ k = m \}
\end{equation}
where $J_{(k)}$ is given by \eqref{eq:J}. It is easily seen that \eqref{eq:jumpval2} reduces to \eqref{eq:jumpval} when $m=n,$ and again assigns special status to the largest observation time $X_{(m)}$.  To the author's knowledge, the extension \eqref{eq:jumpval2} in the case of ties 
appears to be new.

Now, let $\widehat H(t)$ be any right-continuous function such that $\widehat H(t) = 1$ for $t \in [0,X_{(1)}),$
and for $r = 1,\ldots,m,$ that $\widehat H(X_{(r)}-) - \widehat H(X_{(r)}) \geq 0$ and $\widehat H(t-) - \widehat H(t) = 0$ for $t \neq X_{(r)}.$ 
Then, $\widehat H(t)$ is a non-increasing right-continuous step function where jumps can only (but do not always) occur at $X_{(r)}, r = 1,\ldots,m.$ Letting $k_t = \max\{ j: X_{(j)} \leq t\}$ for $t \geq 0$ (i.e., with $X_{(0)} = 0$), 
it follows that
\[
\widehat H(t) = 1- \sum_{r=1}^{k_t} \{ \widehat H(X_{(r)}-) - \widehat H(X_{(r)}) \}.
\]
Because $\widehat S_{\mathit{RTTR}}(t)$ is an example of $\widehat H(t),$ we can therefore represent
\begin{align}
\nonumber
\widehat S_{\mathit{RTTR}}(t) & = 1- \sum_{r=1}^{k_t} 
\{ \widehat S_{\mathit{RTTR}}(X_{(r)}-) - \widehat S_{\mathit{RTTR}}(X_{(r)}) \} \\
\label{eq:rttr-gen}
 & = 1- \sum_{r=1}^{k_t} \left\{
 J_{(r)}  \Delta N(X_{(k)})  I\{ r < m \} + 
J_{(m)} I\{ r = m \} \right\}.
\end{align}
The following proposition extends the known result that $\widehat S_{\mathit{RTTR}}(t) = \widehat S_{\mathit{SC}}(t)$ for $t \geq 0$ to the case of tied data; the results in \citet{Efron1967} for the case of no ties (i.e., $m=n$) are easily obtained as a special case. In addition, this result clarifies the respective connections between $\widehat S_{\mathit{RTTR}}(t)$ and the estimators 
$\widehat F_{\mathit{IPCW}}(t)$ and $\widetilde S_{\mathit{IPCW}}(t)$ defined in the previous section.

\begin{prop}
\label{thm:rttr-ipw-ties}
Suppose $m \leq n$ (i.e.,  ties may exist), and let $\widehat S_{\mathit{RTTR}}(t)$
be the estimator of $S(t)$ obtained using the RTTR algorithm, the jump sizes
being given by \eqref{eq:jumpval2}. Then, $\widehat S_{\mathit{RTTR}}(t) = \widehat S_{\mathit{SC}}(t)$
for $t \geq 0.$ Moreover, defining $\widehat F_{\mathit{RTTR}}(t) = 1 - \widehat S_{\mathit{RTTR}}(t)$,
we have $\widehat F_{\mathit{RTTR}}(t) = \widehat F_{\mathit{IPCW}}(t)$ for $t < X_{(m)}.$ This relationship holds
for $t \geq 0$ if $\Delta N(X_{(m)}) = Y(X_{(m)}),$ in which case $\widehat S_{\mathit{RTTR}}(t) = 
\widetilde S_{\mathit{IPCW}}(t)$ for $t \geq 0.$
\end{prop}

Proposition \ref{thm:rttr-ipw-ties} shows that the relationship between the IPCW and RTTR estimators for the cumulative distribution function $F(t) = 
P\{T \leq t\}$ holds in general provided we consider values of $t$ bounded away from the tail; equivalence for all
$t \geq 0$ only follows under the additional condition that $\Delta N(X_{(m)}) = Y(X_{(m)}).$ 
Interestingly, the result further shows
that $\widehat S_{\mathit{RTTR}}(t)$ does
not have a simple IPCW representation
unless this same tail condition holds;
rather, when $t < X_{(m)},$ the identity \eqref{bigres}
implies 
\[
\widehat S_{\mathit{RTTR}}(t)
=1 - \frac{1}{n}  \sum_{i=1}^n \frac{D_i  }{\widehat{K}(X_i-)} + \widetilde{S}_{\mathit{IPCW}}(t).
\]

The result that $\widehat{S}_{\mathit{RTTR}}(X_{(m)}) = \widehat{S}_{\mathit{SC}}(X_{(m)}) = 0$ together with \eqref{eq:rttr-gen} provides a useful interpretation of the identity \eqref{eq:keysum} in Proposition \ref{thm:surv-ipw}; in particular, the first summand on the left-hand side of \eqref{eq:keysum} represents the total mass placed on observations where failures occur, and the second summand represents the total mass placed on the last observation, or equivalently, the total mass placed on censored observations. This interpretation reflects the fact that the RTTR algorithm assigns  observations $X_{(j)} < X_{(m)}$ positive mass only when there is at least one failure occurring  at $X_{(j)};$ hence, the only possible place where any censored observation(s) can receive mass is at $X_{(m)}.$ 

Although not explicitly recognized in \cite{Efron1967}, Proposition \ref{thm:rttr-ipw-ties} demonstrates that the RTTR estimator
for $F(t)$ is perhaps the first construction of an IPCW estimator, predating the seminal work of \citet{RR1992} by 25 years.  Interestingly, in later work, \citet{Satten2001} characterize the RTTR estimator as an imputation rather than an IPCW estimator, despite the latter arguably providing a more natural characterization.

\section{Discussion}

The results of this paper reconsider several important estimators for the survival function and establish direct connections between them in a novel way. The results presented in Propositions \ref{prop:K2}-\ref{thm:surv-ipw} highlight the central importance of the Volterra integral equation of the first kind, as well as the strong interdependence between the problems of nonparametric estimation for $S(t)$ and $K(t)$. The novel and transparent proofs make use of known theory for Volterra equations, and avoid inductive arguments; a supplementary appendix gives distinct, alternative proofs of Propositions \ref{prop:cdf-ipw} and \ref{thm:surv-ipw} for interested readers.
We also shed new insight on the form of the estimator obtained from Efron's redistribution-to-the-right algorithm, respectively establishing a closed form representation in the presence of ties, conditions under which it can be represented as an IPCW estimator, and a new and simple induction-free proof that this estimator reduces to that generated by the self-consistency equation \eqref{SC1}. 

Self-consistency has been considered as a way to motivate estimators in numerous problems, including problems involving more complex forms of censoring; see, for example, \citet{Turnbull1974, Turnbull1976}, \citet{Mykland1996}, \citet{Huang1996}, and \citet{Mangalam2008}. It would be interesting to investigate the extent to which analogous arguments may help in characterizing the solutions to other problems that satisfy a version of the self-consistency property, and whether there are analogous connections to appropriate Volterra integral equations and/or IPCW-like estimators. \citet{Tsai1985} show in general censoring problems that self-consistent estimators are nonparametric maximum likelihood estimators and hence satisfy certain asymptotic optimality criteria \citep{groeneboom1992information}.  Although self-consistency is neither the most convenient nor contemporary approach to estimator computation, it remains an interesting and useful approach to characterizing estimators that may have good efficiency properties. 

\section*{Acknowledgement}

This revision of an earlier work due to \citet{RLS2023} was initiated during BRB's postdoctoral study at the University of Rochester. RLS and BRB contributed equally to its final content. RLS and BRB thank David Oakes for a helpful conversation that led to the supplemental proof of Proposition \ref{prop:cdf-ipw}; BRB also thanks David Oakes for an introduction to the redistribute-to-the-right algorithm.

\newpage

\bibliographystyle{apalike}
\bibliography{master_bib}

\newpage
\appendix

\section*{Appendix}

\subsection*{Proof of Proposition \ref{prop:K2}}
\begin{proof}
To prove \eqref{PL-gen-K}, we begin by noting that $t < X_{(m)}$ guarantees that
$Y(u+) > 0$ for $u \leq t$. Hence, using results briefly summarized in \citet[p. 1526]{Gill1990} for general forms of anticipating Volterra integral equations, the unique solution to \eqref{VoltK} is  given by 
    \begin{equation}
    \label{for-rr-later}
        \widehat{K}(t)
        = \left[ \Prodi_{(0,t]} \left\{ 1 + \frac{\d C(u)}{Y(u+)} \right\} \right]^{-1}
        = \left[ \prod_{(0,t]} \left\{ 1 + \frac{\Delta C(u)}{Y(u+)} \right\} \right]^{-1}.
    \end{equation}
    Here, the finite product form given in the last equality on the right-hand side arises because $C(\cdot)$ can only jump at a finite set of times. To see that \eqref{for-rr-later} reduces to \eqref{PL-gen-K}, observe that straightforward algebra yields
    \begin{equation*}
        1/\left\{ 1 + \frac{\Delta C(u)}{Y(u+)} \right\} 
        = \frac{Y(u+)}{Y(u+) + \Delta C(u)} 
        =  1 - \frac{\Delta C(u)}{Y(u+) + \Delta C(u)}.
    \end{equation*}
    Now, we can write $Y(u)-Y(u+) = \Delta N(u) + \Delta C(u);$
    hence, $Y(u+) + \Delta C(u) = Y(u) - \Delta N(u) = Y^{\dagger}(u)$ and it follows immediately that 
     \begin{equation}
     \label{anticipatePL}
       \left[ \prod_{(0,t]} \left\{ 1 + \frac{\Delta C(u)}{Y(u+)} \right\} \right]^{-1}
        = 
         \prod_{(0,t]} \left\{ 1 - \frac{\Delta C(u)}{Y^{\dagger}(u)} \right\} = 
        \Prodi_{(0,t]} \left\{ 1 - \frac{\d C(u)}{Y^{\dagger}(u)} \right\},
         \end{equation}
establishing \eqref{PL-gen-K}. 

To see that result \eqref{K-PL} follows under the assumption that $\Delta N(u) 
\Delta C(u) = 0$ for $u< X_{(m)}$, we simply note
\begin{equation*}
 \widehat{K}(t) = 
         \prod_{(0,t]} \left\{ 1 - \frac{\Delta C(u)}{Y^{\dagger}(u)} \right\} =
         \prod_{(0,t]} \left\{ 1 - \frac{\Delta C(u)}{Y(u)-
         \Delta N(u)} \right\}
         = \prod_{(0,t]} \left\{ 1 - \frac{\Delta C(u)}{Y(u)} \right\},
\end{equation*}
which is equivalent to \eqref{K-PL}. This latter form is also the known unique solution
to the integral equation for $\widehat K(t)$ stated in the theorem for
the case where $\Delta N(u) \Delta C(u) = 0$ for $u< X_{(m)};$
see, for example, Theorem II.6.1 in \citet[Sec.\ II.6]{ABGK1993}.
\end{proof}

\subsection*{Proof of Proposition \ref{prop:PL-KM}}
\begin{proof}
The result \eqref{eq:fail-prod-lim-SC} follows immediately for $t = 0$ and $t \geq X_{(m)}$ by previous arguments; hence, we must only show that $\widehat S_{\mathit{SC}}(t)$ equals $\widehat S_{\mathit{PL}}(t)$ given in \eqref{eq:fail-prod-lim} for $t \in (0,X_{(m)}).$

Suppose $\widehat S(t)$ solves \eqref{SC2}, equivalently \eqref{SC1}, for $t \in (0,X_{(m)}).$  Proposition \ref{prop:K2} and
the definition of $\widehat K(t)$ in \eqref{Khat} then imply that the solution $\widehat S(t)$ to \eqref{SC2} necessarily must also satisfy
      \begin{equation*}
    \widehat{S}(t) \times
         \prod_{(0,t]} \left\{ 1 - \frac{\Delta C(u)}{Y^{\dagger}(u)} \right\}
         = \frac{Y(t+)}{n}.
  \end{equation*}
    The crux of the proof thus involves determining a suitable relationship between $n^{-1} Y(t+)$ and 
    the product-limit form of $\widehat{K}(t)$ in Proposition \ref{prop:K2}.
    
    We begin by finding a comparable finite-product representation for $n^{-1} Y(t+)$. Using a telescoping product and the fact that $Y(0+) = n$, we can write
    \begin{equation*}
        \frac{Y(t+)}{n}
        = \frac{Y(t+)}{Y(0+)}
        = \prod_{0 < u \leq t} \frac{Y(u+)}{Y(u)}
        = \prod_{0 < u \leq t} \left\{ 1 - \frac{\Delta N(u)}{Y(u)} - \frac{\Delta C(u)}{Y(u)} \right\}, 
    \end{equation*}
    where the last equality follows from $Y(u+) = Y(u) - \Delta N(u) - \Delta C(u)$. 
    Now, since $Y^{\dagger}(u) = Y(u) - \Delta N(u),$ observe that
    \begin{equation*}
        \frac{\Delta C(u)}{Y(u)}
        = \frac{Y^{\dagger}(u)}{Y(u)} \frac{\Delta C(u)}{Y^{\dagger}(u)} 
        = \left\{ 1 - \frac{\Delta N(u)}{Y(u)} \right\} \frac{\Delta C(u)}{Y^{\dagger}(u)};
   \end{equation*}
   hence, 
    \begin{equation*}
        1 - \frac{\Delta N(u)}{Y(u)} - \frac{\Delta C(u)}{Y(u)}
        = \left\{ 1 - \frac{\Delta N(u)}{Y(u)} \right\} \left\{ 1 -  \frac{\Delta C(u)}{Y^{\dagger}(u)} \right\}. 
    \end{equation*}
   Taking the product of both sides over all jumps, we see that 
   \begin{align}
        \frac{Y(t+)}{n} 
        & = \prod_{(0,t]} \left\{ 1 - \frac{\Delta N(u)}{Y(u)} \right\} \left\{ 1 -  \frac{\Delta C(u)}{Y^{\dagger}(u)} \right\} \nonumber \\
        & = \prod_{(0,t]}\left\{ 1 - \frac{\Delta N(u)}{Y(u)} \right\} \times \prod_{(0,t]} \left\{ 1 -  \frac{\Delta C(u)}{Y^{\dagger}(u)} \right\} \nonumber \\
        & = \Prodi_{(0,t]} \left\{ 1 - \frac{\d N(u)}{Y(u)} \right\} \times \widehat{K}(t). \label{Gill-ident}
   \end{align}
   Therefore, since $\widehat K(t) \widehat S(t) = n^{-1} Y(t+),$ we have
    \[
    \widehat{S}(t) = 
    \frac{n^{-1} Y(t+)}{\widehat{K}(t)}
    = \Prodi_{(0,t]} \left\{ 1 - \frac{\d N(u)}{Y(u)}
    \right\} = \widehat{S}_{\mathit{PL}}(t),
    \]
    the latter being given in 
    \eqref{eq:fail-prod-lim} and proving
    the desired result. We note here
    that the relationship derived in 
   \eqref{Gill-ident} is that 
   originally given in \citet[p. 36]{Gill1980}; to see this, note that
   (i) $Y^{\dagger}(t) = Y(t) - \Delta N(t);$ and,
   (ii) 
    $Y(t+) = Y(t) - \Delta N(t) - \Delta C(t)$
    gives $\Delta C(t) = Y(t) - Y(t+) - \Delta N(t).$
\end{proof}

\subsection*{Proof of Proposition \ref{prop:cdf-ipw}}

\begin{proof}
The result is trivial for $t =0$. 
For $t > 0,$ previous results imply that the relation \eqref{Khat} can be 
rewritten as $\widehat{K}(u) \widehat{S}_{\mathit{PL}}(u) = Y(u+)/n;$ 
as seen in \eqref{Gill-ident}, this relation is in fact valid 
for $u > 0.$  Taking the limit of this equation from the left,
it follows that $\widehat{K}(u-) \widehat{S}_{\mathit{PL}}(u-) = Y(u)/n$ for $u > 0$. Since $Y(u) > 0$ for $u \leq X_{(m)}$, 
we have $\widehat{K}(u-) > 0$ and $\widehat{S}_{\mathit{PL}}(u-) > 0$ for $u \leq X_{(m)}$. Substitution of 
$\widehat{S}_{\mathit{PL}}(u-) = n^{-1} Y(u) / \widehat{K}(u-)$ into the right-hand side of the integral equation \eqref{VoltS} then gives
    \begin{equation*}
        \widehat{F}_{\mathit{PL}}(t) =
        1 - \widehat{S}_{\mathit{PL}}(t) 
        = \frac{1}{n}  \int_{(0,t]}  \frac{\d N(u)}{\widehat{K}(u-)}.
    \end{equation*}
    This last expression remains 
    well-defined for $t > X_{(m)}$ 
    since $\widehat{K}(X_{(m)}-) > 0$ and $\Delta N(u) = 0$ for $u > X_{(m)}$. 
    The desired result now follows immediately upon
    evaluating the integral. 
\end{proof}

\subsection*{Proof of Proposition \ref{thm:surv-ipw}}
\begin{proof}

The proof of Proposition \ref{prop:cdf-ipw} and
\eqref{bigres} imply
\begin{equation*}
\widehat{S}_{\mathit{PL}}(X_{(m)}) =
1 - \frac{1}{n} \sum_{i=1}^n \frac{D_i  }{\widehat{K}(X_i-)} - \widetilde{S}_{\mathit{IPCW}}(X_{(m)});
\end{equation*}
since $X_{(m)}$ is the largest observation time
and $\widehat{K}(X_i-) >0$ for each $i,$
it is trivially true that $ \widetilde{S}_{\mathit{IPCW}}(X_{(m)}) = 0.$
Consequently,
\begin{equation*}
\widehat{S}_{\mathit{PL}}(X_{(m)}) =
1 - \frac{1}{n} \sum_{i=1}^n \frac{D_i  }{\widehat{K}(X_i-)},
\end{equation*}
and \eqref{eq:keysum} therefore holds if and only if 
\[
\widehat{S}_{\mathit{PL}}(X_{(m)}) = \frac{\Delta C(X_{(m)})}{n \widehat{K}(X_{(m)}-)}.
\]
Towards this end, note that
\[
\widehat{S}_{\mathit{PL}}(X_{(m)}) = \widehat{S}_{\mathit{PL}}(X_{(m)}-)
\left(1 - \frac{\Delta N(X_{(m)})}{Y(X_{(m)})} \right)
= \widehat{S}_{\mathit{PL}}(X_{(m)}-) \frac{\Delta C(X_{(m)})}{Y(X_{(m)})},
\]
the last equality following from the fact that $Y(X_{(m)}) = \Delta N(X_{(m)}) + \Delta C(X_{(m)})$. Since $\widehat{K}(u-) \widehat{S}_{\mathit{PL}}(u-) = Y(u)/n$ for $u > 0,$ we have for $u = X_{(m)}$ that
\[
 \frac{\widehat{S}_{\mathit{PL}}(X_{(m)}-)}{Y(X_{(m)})} = \frac{1}{n \widehat{K}(X_{(m)}-)},
\]
proving \eqref{eq:keysum}. In view of the fact that
$Y(X_{(m)}) = \Delta N(X_{(m)}) + \Delta C(X_{(m)}),$
we also see that \eqref{eq:keysum2} holds 
if and only if $\Delta C(X_{(m)}) = 0$
$\Leftrightarrow$ 
$Y(X_{(m)}) = \Delta N(X_{(m)}).$ Finally, 
it is immediate from \eqref{bigres} that
$\widetilde{S}_{\mathit{IPCW}}(t) = 1- \widehat{F}_{\mathit{IPCW}}(t)$ 
for $t \geq 0$ if and only if \eqref{eq:keysum2} holds.
\end{proof}

\subsection*{Proof of Proposition \ref{thm:rttr-ipw-ties}}

\begin{proof}
Because the RTTR algorithm starts with a total mass of one,
the preservation of mass through redistribution implies
that
\begin{equation}
\label{eq:jumpsum}
\sum_{k=1}^{m-1} J_{(k)} \Delta N(X_{(k)}) + 
Y(X_{(m)})  J_{(m)}  = 1.
\end{equation}
Hence, using \eqref{eq:rttr-gen}, we find that
\begin{eqnarray*}
\widehat S_{\mathit{RTTR}}(t) & = & 1- \sum_{r=1}^{k_t} \left( 
 J_{(r)} \Delta N(X_{(r)}) I\{ r < m \} + 
J_{(m)} Y(X_{(m)}) I\{ r = m \} \right) \\
& = &
\begin{cases}
1 - \sum_{r=1}^{m-1} 
 J_{(r)} \Delta N(X_{(r)}) - J_{(m)} Y(X_{(m)}) & t \geq X_{(m)} \\
 1 - \sum_{r=1}^{k_t} 
 J_{(r)} \Delta N(X_{(r)}) & t < X_{(m)}
\end{cases} \\
& = &
\begin{cases}
0  & t \geq X_{(m)} \\
 1 - \sum_{r=1}^{k_t} 
 \frac{\Delta N(X_{(r)})}{n \widehat K(X_{(r)}-)}& t < X_{(m)}
\end{cases},
\end{eqnarray*}
the last representation respectively following from \eqref{eq:jumpsum} and \eqref{eq:jumpval2}. Focusing on the case where $t < X_{(m)},$ it is easily seen that
\[
\sum_{r=1}^{k_t} 
 J_{(r)} \Delta N(X_{(r)})= \sum_{i=1}^n I\{ X_i \leq t \} 
 \frac{D_i}{n \widehat K(X_i-)} = \widehat F_{\mathit{IPCW}}(t),
\]
where $\widehat F_{\mathit{IPCW}}(t)$ is defined in \eqref{eq:ipcw-defs}. 
Therefore,
\[
\widehat S_{\mathit{RTTR}}(t)  =  
\begin{cases}
0  & t \geq X_{(m)} \\
 1 - \widehat F_{\mathit{IPCW}}(t) & t < X_{(m)}
\end{cases} \, = \,
\begin{cases}
0  & t \geq X_{(m)} \\
\widehat S_{\mathit{PL}}(t) & t < X_{(m)}
\end{cases},
\]
the last equality following by Proposition \ref{prop:cdf-ipw}
and \eqref{bigres}. The result \eqref{eq:fail-prod-lim-SC} in Proposition \ref{prop:PL-KM}
now establishes the desired equivalence, that is, $\widehat S_{\mathit{RTTR}}(t) = \widehat S_{\mathit{SC}}(t)$
for $t \geq 0.$ 

Moreover, these same calculations prove that $\widehat F_{\mathit{RTTR}}(t) = 1 - \widehat S_{\mathit{RTTR}}(t) =\widehat F_{\mathit{IPCW}}(t)$ for $t < X_{(m)};$ by Proposition \ref{thm:surv-ipw}, this relationship between the IPCW and RTTR estimator holds for $t \geq 0$ under the additional condition that $\Delta N(X_{(m)}) = Y(X_{(m)})$ (i.e., all observations at the last observation time are failures), in which case $\widehat S_{\mathit{RTTR}}(t) = \widetilde S_{\mathit{IPCW}}(t)$ for $t \geq 0.$

The arguments above are predicated on the validity of \eqref{eq:jumpsum}; we now provide independent verification that \eqref{eq:jumpsum} holds. Recalling the fact that $Y(X_{(m)}) = \Delta N(X_{(m)}) + \Delta C(X_{(m)}),$ we can rewrite the left-hand side of \eqref{eq:jumpsum} as
\[
\sum_{k=1}^{m} J_{(k)} \Delta N(X_{(k)}) + \Delta C(X_{(m)}) J_{(m)}.
\]
Using the definition of $J_{(k)},$ it is easy to show that
\[
\sum_{k=1}^{m} J_{(k)} \Delta N(X_{(k)}) = 
\frac{1}{n} \sum_{i=1}^n \frac{D_i}{\widehat{K}(X_i-)}
\]
and 
\[
\Delta C(X_{(m)}) J_{(m)} = 
\frac{\Delta C(X_{(m)})}{n \widehat K(X_{(m)}-)};
\]
hence, the left-hand side of \eqref{eq:jumpsum} reduces
to
\[
\frac{1}{n} \sum_{i=1}^n \frac{D_i}{\widehat{K}(X_i-)} 
+ \frac{\Delta C(X_{(m)})}{n \widehat K(X_{(m)}-)},
\]
which is indeed equal to one by \eqref{eq:keysum}.
\end{proof}

\newpage
\section*{Supplementary Appendix}
In this supplementary appendix, we respectively provide alternative proofs of Propositions \ref{prop:cdf-ipw} and \ref{thm:surv-ipw}
that may be of independent interest to readers.

\subsection*{Alternative proof of Proposition \ref{prop:cdf-ipw}}

The following proof uses a different approach than that given in the main Appendix,  and represents a minor generalization of a proof communicated by David Oakes in the case that $\Delta N(\cdot) \Delta C(\cdot) = 0$. Under this condition, the proof below can be considerably simplified and made more transparent by only considering $t>0$ such that $\Delta  \widehat{S}_{\mathit{PL}}(t)  = \widehat{S}_{\mathit{PL}}(t-) - \widehat{S}_{\mathit{PL}}(t) > 0$. 

\begin{proof}
Consider $t < X_{(m)}$. We may see that 
\begin{equation}
    \Delta \left\{ \frac{Y(t+)}{n} \right\} 
    = \frac{\Delta N(t) + \Delta C(t)}{n} \label{eq:oakes1}
\end{equation}
and 
\begin{align}
    \Delta \left\{ \widehat{S}_{\mathit{PL}}(t) \widehat{K}(t) \right\}
    & = \Delta \left\{ \widehat{S}_{\mathit{PL}}(t)  \right\} \widehat{K}(t) + \widehat{S}_{\mathit{PL}}(t-) \Delta \left\{ \widehat{K}(t) \right\} \label{eq:oakes2} \\
    & = \Delta \left\{ \widehat{S}_{\mathit{PL}}(t)  \right\} \widehat{K}(t-) \left\{ 1 - \frac{\Delta C(t)}{Y^{\dagger}(t)} \right\} + \widehat{S}_{\mathit{PL}}(t-) \widehat{K}(t-) \frac{\Delta C(t)}{Y^{\dagger}(t)} \nonumber \\
    & = \Delta \left\{ \widehat{S}_{\mathit{PL}}(t)  \right\} \widehat{K}(t-)\frac{Y(t+)}{Y^{\dagger}(t)} + \frac{Y(t)}{n} \frac{\Delta C(t)}{Y^{\dagger}(t)}. \nonumber
\end{align}

Since $Y(u+) / n = \widehat{S}_{\mathit{PL}}(u) \widehat{K}(u)$ for all $u>0$, the
increments $n^{-1} \Delta Y(t+)$ and $\Delta \left\{ \widehat{S}_{\mathit{PL}}(t) \widehat{K}(t) \right\}$ must be equal. Equating these and rearranging terms, we have derived that 
\begin{align*}
    n \Delta \left\{ \widehat{S}_{\mathit{PL}}(t)  \right\} \widehat{K}(t-)
    & = \frac{Y^{\dagger}(t)}{Y(t+)} \left\{ \Delta N(t) + \Delta C(t) - \frac{Y(t)}{Y^{\dagger}(t)} \Delta C(t) \right\} \\
    & = \frac{Y^{\dagger}(t)}{Y(t+)} \left\{ \Delta N(t) - \frac{\Delta C(t) \Delta N(t)}{Y^{\dagger}(t)}  \right\} \\
    & = \Delta N(t) \frac{Y^{\dagger}(t)}{Y(t+)} \left\{ 1 - \frac{\Delta C(t)}{Y^{\dagger}(t)}  \right\} \\
    & = \Delta N(t). 
\end{align*}
Therefore 
\begin{align*}
    \widehat{S}_{\mathit{PL}}(t)
    & = 1 - \sum_{u \leq t \, : \, \Delta N(u) > 0} \Delta \left\{ \widehat{S}_{\mathit{PL}}(u)  \right\} 
    = 1 - \sum_{u \leq t \, : \, \Delta N(u) > 0} \frac{\Delta N(u)}{n \widehat{K}(u-)}
    = 1 - \frac{1}{n} \int_{(0,t]}  \frac{\d N(u)}{\widehat{K}(u-)}.
\end{align*}
Evaluating the integral as in the main proof, we obtain $ \widehat{S}_{\mathit{PL}}(t)=1-\widehat{F}_{\mathit{IPCW}}(t)$, as desired. The behavior for $t \geq X_{(m)}$ follows as in the previous proof. 
\end{proof}

We now show how the proof can be simplified in the special case that  $\Delta N(\cdot) \Delta C(\cdot) = 0$. Consider $t \in (0, X_{(m)})$ such that $\Delta  \widehat{S}_{\mathit{PL}}(t)  = \widehat{S}_{\mathit{PL}}(t-) - \widehat{S}_{\mathit{PL}}(t) > 0$. Then it must be that $\Delta N(t) > 0$ and hence that $\Delta C(t) = 0$. 
The first equation, \eqref{eq:oakes1}, simplifies to
\begin{equation*}
    \Delta \left\{ \frac{Y(t+)}{n} \right\} 
    = \frac{\Delta N(t)}{n} 
\end{equation*}
and the second equation, \eqref{eq:oakes2}, simplifies to
\begin{equation*}
    \Delta \left\{ \widehat{S}_{\mathit{PL}}(t) \widehat{K}(t) \right\}
    = \Delta \left\{ \widehat{S}_{\mathit{PL}}(t)  \right\} \widehat{K}(t),
\end{equation*}
since $\Delta \left\{ \widehat{K}(t) \right\}=\widehat{K}(t-) \Delta C(t) / Y^{\dagger}(t) = 0$. Notice that no calculation was required for either of these equalities. The conclusion that $n \Delta \left\{ \widehat{S}_{\mathit{PL}}(t)  \right\} \widehat{K}(t-) = \Delta N(t)$ follows immediately.

\subsection*{Alternative proof of Proposition \ref{thm:surv-ipw}}
    
Below, we present an alternative proof of Proposition \ref{thm:surv-ipw},
specifically the derivation of conditions under which \eqref{eq:keysum2}
holds, as well as the identity \eqref{eq:keysum}. 
    
\begin{proof}
    Let $X_{(f)} \leq X_{(m)}$ be the largest observation time such that there is at least one observation $X_i$ with $D_i = 1;$ substitution of $t = X_{(f)}$ into \eqref{bigres} gives
    \begin{equation*}
        \widehat{S}_{\mathit{PL}}(X_{(f)}) =
        1 - \frac{1}{n} \sum_{i=1}^n \frac{D_i  }{\widehat{K}(X_i-)} - \widetilde{S}_{\mathit{IPCW}}(X_{(f)}).
    \end{equation*}
    Observe that the definition of $X_{(f)}$ implies
    \[
    \widetilde{S}_{\mathit{IPCW}}(X_{(f)}) =
    \frac{1}{n} \sum_{i=1}^n \frac{D_i I \{X_i > X_{(f)} \} }{\widehat{K}(X_i-)}
    = 0
    \]
    since $\widehat{K}(X_i-) >0$ and
    $D_i I \{X_i > X_{(f)} \} = 0$ for $i=1,\ldots,n.$ 
    Consequently,
       \begin{equation*}
        \widehat{S}_{\mathit{PL}}(X_{(f)}) =
        1 - \frac{1}{n} \sum_{i=1}^n \frac{D_i  }{\widehat{K}(X_i-)}.
        \end{equation*}
    The last equality above clearly implies \eqref{eq:keysum2} holds if and only if $\widehat{S}_{\mathit{PL}}(X_{(f)}) = 0.$ Since
    \begin{equation*}
        \widehat{S}_{\mathit{PL}}(X_{(f)})
        = \left[ \Prodi_{(0,X_{(f)})} \left\{ 1 - \frac{\d N(u)}{Y(u)} \right\} \right] \left\{ 1 - \frac{\Delta N(X_{(f)})}{Y(X_{(f)})} \right\}
    \end{equation*}
    and the first term in the product on the right-hand side never vanishes, it follows that $\widehat{S}_{\mathit{PL}}(X_{(f)}) = 0$ if and only if 
    $\Delta N(X_{(f)}) = Y(X_{(f)}).$ However, such an equality can only 
    occur when (i) $X_{(f)} = X_{(m)}$ {\em and} (ii) that 
    all observations with $X_i = X_{(m)}$ have $D_i = 1.$  In general,
    $X_{(f)} = X_{(m)}$ is necessary, but not sufficient, for $\Delta N(X_{(m}) = Y(X_{(m)})$ since both failure and censoring can occur at $X_{(m)}.$

    To establish \eqref{eq:keysum}, we first note that 
    $\widehat{S}_{\mathit{PL}}(X_{(f)}) = \widehat{S}_{\mathit{PL}}(X_{(m)}).$
    This result is immediate when $f = m,$ so consider $f < m.$
    In this case, by definition of $X_{(f)},$ we must
    have $\Delta N(X_{(j)}) = 0$ for $j = f + j,$ $j = 1, \ldots, m-f.$
    Consequently, by considering the behavior of the product integral representation for $\widehat S_{\mathit{PL}}(\cdot),$ it is easily shown
    that $\widehat{S}_{\mathit{PL}}(X_{(f)}) = \widehat{S}_{\mathit{PL}}(X_{(m)}).$ Establishing
    \eqref{eq:keysum} now follows as in the main proof; in particular, since
    \[
    \widehat{S}_{\mathit{PL}}(X_{(m)}) = \frac{\Delta C(X_{(m)})}{n \widehat{K}(X_{(m)}-)}
    \]
    and $\widehat{S}_{\mathit{PL}}(X_{(f)}) = \widehat{S}_{\mathit{PL}}(X_{(m)}),$
    the identity \eqref{eq:keysum} must hold. That
    $\widetilde{S}_{\mathit{IPCW}}(t) = 1- \widehat{F}_{\mathit{IPCW}}(t)$ 
    for $t \geq 0$ if and only if \eqref{eq:keysum2} holds
    also follows as in the main proof.
    \end{proof}

\end{document}